\documentclass[%
 reprint,
 amsmath,amssymb,
 aps,
]{revtex4-2}

\usepackage{graphicx}% Include figure files
\usepackage{dcolumn}% Align table columns on decimal point
\usepackage{bm}% bold math
\usepackage{natbib}
\begin{document}

\preprint{APS/123-QED}

\title{Multi-Messenger Constraint on the Hubble Constant $H_0$ \\
with Tidal Disruption Events}% Force line breaks with \\

\author{Thomas Hong Tsun Wong}
 \altaffiliation[Email: ]{h7wong@ucsd.edu}%Lines break automatically or can be forced with \\
\affiliation{%
 Department of Physics, University of California, San Diego, California, 92092, USA}%

\date{\today}% It is always \today, today,

\begin{abstract}
Tidal disruption events (TDEs), apart from producing luminous electromagnetic (EM) flares, can generate potentially detectable gravitational wave (GW) burst signals by future space-borne GW detectors.  In this Letter, we propose a methodology to constrain the Hubble constant $H_0$ by incorporating the TDE parameters measured by EM observations (e.g., stellar mass, black hole (BH) mass and spin, and other orbital parameters) into the observed TDE GW waveforms.  We argue that an accurate knowledge of the BH spin could help constrain the orbital inclination angle, hence alleviating the well-known distance-inclination degeneracy in GW waveform fitting.  For individual TDEs, the precise redshift measurement of the host galaxies along with the luminosity distance $D_{\rm L}$ constrained by EM and GW signals would give a self-contained measurement of $H_0$ via Hubble's law, completely independent of any specific cosmological models. 
\end{abstract}

% Read Lars' Skype comments: 
% 1. Cannot try to model an actual event and see how well we overcome degeneracy
% 2. Different ways to infer TDE parameters assume widely different geometry of emission -> different TDE geometry
% 3. Way to show method works: Adopt some typical uncertainties (PDF) for each parameter and use them as prior to fit in GW waveform, compared with no prior, then see if H0 is more precise/more accurate.  Once D_L uncertainty decreases by incorporating EM constraints, then this method works. 
% 4. Could assume linear interpolation for GW catalog to sample the parameter space
% 5. Keep the not-well-constrained parameters constant and fit the inclination, MBH, beta~
% 6. Fe K alpha line only from one TDE source, extremely rare to find, X-ray reflection/reverberation is rare
% 7. Require computationally more advanced way to track the exact stream multiple windings, and *how the self-intersection results in the specific accretion disk plane*

\maketitle

%\tableofcontents

\section{Introduction} \label{sec:intro}
%The local expansion rate of the Universe can be described by Hubble's law: 
%\begin{equation} \label{eq:Hubble}
%    v_{\rm H} = cz = H_0 D_{\rm L}~,
%\end{equation}
%where the Hubble constant $H_0$ measures the recessional velocity $v_{\rm H}$ of galaxies located at a certain luminosity distance $D_{\rm L}$ away from us, which scales linearly with redshift $z$.  This constant has fundamental importance as it sets the overall size of the Universe.  In search of the precise value for $H_0$, independent observations of the early Universe (e.g., cosmic microwave background) and the late-time observables (e.g., Type Ia supernova as ``standard candles'') are applied onto the widely acknowledged cosmological $\Lambda$CDM model \cite{Verde+19}. 

The method of utilizing the emissions of gravitational wave (GW) by compact object mergers, known as the ``standard sirens'' (well-defined sources emitting at some known frequencies), to measure $H_0$ was long proposed \cite{Schutz86}.  From Hubble's law: 
\begin{equation} \label{eq:Hubble}
    v_{\rm H} = cz = H_0 D_{\rm L}~,
\end{equation}
the detected GW waveform provides constraints on $D_{\rm L}$ while the bright electromagnetic (EM) counterpart measures the redshift, known as the ``bright siren'' (an EM-observable ``standard siren'').  It was only until recently has this technique been implemented in the binary neutron star merger event GW170817 \cite{Abbott+17}.  The uncertainty in $H_0$ measurement is, however, dominated by the degeneracy between $D_{\rm L}$ and the inclination angle $\eta$, defined here as the angle between the orbital angular momentum and the line of sight \footnote{Note that this definition differs by $90^\circ$ as compared with the conventional description}, of the binary system from the GW template-waveform fitting \cite{Abbott+17, Bulla+22}, as seen for a small-angle approximation: 
\begin{equation} \label{eq:degeneracy}
    h_{\rm GW} \propto \frac{\cos\eta}{D_{\rm L}}~,
\end{equation}
where $h_{\rm GW}$ is the detected GW strain amplitude.  By incorporating the multi-messenger information of the event, the viewing-angle-dependent features of various EM emission models (e.g., gamma-ray burst and kilonova) are exploited to arbitrate the distance-inclination degeneracy, providing a tighter constraint on $D_{\rm L}$, and subsequently $H_0$ (\cite{Bulla+22} and references therein).  

One would naturally question whether compact object mergers remain the sole astrophysical sources to measure $H_0$ in a multi-messenger approach.  As long as massive objects revolve around each other, GW emissions are guaranteed, therefore tidal disruption of stars by massive black holes (BHs) would present themselves as viable candidates due to the fact that immense EM radiation is released during the transient event \cite{Kobayashi+04, Eracleous+19, Pfister+22, Toscani+22}.  When a star approaches the galactic central supermassive black hole (SMBH) at a sufficiently close distance, the tidal radius $r_{\rm T} \approx \left(M_{\rm BH}/m_\star\right)^{1/3}r_\star$, where $M_{\rm BH}$, $m_\star$, $r_\star$ are the BH mass, stellar mass, and stellar radius, respectively, the star is then torn apart given that the tidal field of the hole exceeds the star's self-gravity \cite{Hills75}.  The disrupted stellar material would be stretched into a debris stream, approximately half of it will gradually dissipate orbital energy into EM radiation, and eventually circularize into an accretion disk.  Optical, near-UV, X-ray all-sky surveys have detected up to a hundred or so events, and one to two orders of magnitude more are expected in the coming decade \cite{French+20, Gezari21}.  

Tidal disruption events (TDEs) could only generate GW bursts as the star is often disrupted within an orbital timescale, i.e. the intact star does not survive an entire orbit to produce a full period of GW waveform \cite{Kobayashi+04}.  An open comprehensive living catalog of TDE GW waveforms has been built to explore a wide range of parameters \cite{Toscani+22}.  Given their relatively long orbital timescale prior to disruption ($\sim10^{2-4}\,{\rm s}$), the characteristic GW burst frequency is approximately in the range of $0.1-10\,{\rm mHz}$, which corresponds to the designed sensitivities of the upcoming space-borne GW detectors \cite{LISA17, TianQin16, DECIGO06, BBO06}.  But in fact, most TDE GW signals are incapable of generating a large enough signal-to-noise ratio to trigger a detection for \emph{LISA} \cite{LISA17} but would lie well within the detection limit of post-\emph{LISA} detectors \cite{DECIGO06, BBO06, Pfister+22, Toscani+23}.  The TDE GW observed rate by \emph{LISA} is predicted to remain half a dozen or so for the entire four-year mission \cite{Pfister+22} as the characteristic strains of the events are weak given typical TDE parameters, which are estimated as \cite{Kobayashi+04, Toscani+22}: 
\begin{equation}\label{eq:h_GW}
\begin{split}
    h_{\rm GW} &\sim 10^{-22}\left(\frac{D_{\rm L}}{20\,{\rm Mpc}}\right)^{-1}\times \\
    &~~~~\,\beta\left(\frac{r_\star}{{\rm R}_\odot}\right)^{-1}\left(\frac{m_\star}{{\rm M}_\odot}\right)^{4/3}\left(\frac{M_{\rm BH}}{10^6\,{\rm M}_\odot}\right)^{2/3}~,
\end{split}
\end{equation}
where $r_{\rm p}$ is the pericenter radius and $\beta=r_{\rm p}/r_{\rm T}$ is the penetration parameter \footnote{As a simple approximation, $\beta\geq1$ usually refers to the complete disruption of the star, but \cite{Guillochon+Enrico13} has shown that the cutoff between survival and disrupted sometimes lies above unity.} (quantifying how deeply the orbit penetrates into the BH gravitational potential well).  

The pessimistic observed rate by \emph{LISA} inspires us to explore the possibility of using a very limited number of events to independently measure $H_0$.  We hereby propose using TDE EM observations to constrain as many parameters as possible prior to GW waveform fitting, resulting in a remarkably improved GW constraint on the luminosity distance $D_{\rm L}$.  This work focuses on gathering the cumulative modeling effort in obtaining TDE parameters, exploring their intercorrelations, and most importantly, proposing the \emph{first} methodology to constrain $H_0$ via TDEs.  Prior to detecting GW signals, using simulated waveforms to run the following analysis in an attempt to constrain $H_0$ would inevitably lead to a circular argument, therefore this letter serves as a primal investigation on the foundational idea of exploiting multi-messenger signals of TDEs to measure $H_0$. 

In Section \ref{sec:DL_constraint}, we illustrate the recent progress in constraining all waveform-dependent TDE parameters by EM observations with physical modeling, laying the foundational work to further propose the novel approach to alleviate the distance-inclination degeneracy using the constrained BH spin parameters, the methodology is then presented with an estimation on which parameter(s) would most dominate the $H_0$ uncertainty.  In Section \ref{sec:discussion}, we discuss possible ways to further improve the precision of $H_0$ determination.

\section{Multi-Messenger Constraints on $D_{\rm L}$} \label{sec:DL_constraint}

Given the ability to localize TDEs with the current multi-wavelength surveys, the redshifts of the host galaxies can be comfortably measured with high certainty \cite{French+20, Gezari21}.  In order to estimate an accurate Hubble constant $H_0$, the problem lies in constraining the luminosity distance $D_{\rm L}$ from their GW counterparts.  

Unfortunately, in order to accurately match the observed waveforms with the templates, one could not rely solely on the approximate peak amplitude in Eq.(\ref{eq:h_GW}) which depends mainly on three parameters.  Provided the number of waveform-dependent parameters, there is only hope to constrain $D_{\rm L}$, and hence $H_0$, if most, if not all, these TDE parameters could be constrained to some extent with EM observations.  We hereby show the parameter inter-dependencies (summarized in Fig.\ref{fig:bayesian}) and how they are expected to yield a constrained $H_0$.

\subsection{EM Constraints on TDE Parameters} \label{subsec:EM_constraint} 
\subsubsection{Three main parameters: $M_{\rm BH}$, $m_\star$, $\beta$} \label{subsubsec:mainparameter}
TDE-host black hole mass is one of the easiest to infer since there are a few $M_{\rm BH}$-galaxy relations available \cite{McConnell+Ma13, Gultekin+19, Reines+Volonteri15} and simulations/models specifically for TDE-specific scenarios \cite{Kobayashi+04, Guillochon+Enrico13, Mockler+19, Ryu+20, Wen+20, Zhou+21}.  The common TDE EM-GW detection rate is highly limited \cite{Pfister+22}, a more accurate constraint on the $M_{\rm BH}$ of the few detectable events could perhaps be computed in a case-by-case TDE-model-dependent manner instead of using the global galaxy relations, even if the latter has a slightly better constraint than the former.  

Focusing on the disruption of main-sequence (MS) stars, the mass-radius relation is given by \cite{Kippenhahn+13}, such that all $r_\star$-dependence turns into $m_\star$ accordingly.  MCMC method that fits both the peak luminosity and the color temperature of the observed TDEs could constrain $m_\star$ down to a $\sim$few \% uncertainty (but sometimes a lot higher) \cite{Ryu+20, Zhou+21}.  It is indeed a challenging task to check whether the mass constraint is accurate given there is yet to exist another independent EM-measurement, additional GW waveform information could complement the deficiency based on the approximate duration $\tau$/frequency $f$ of the burst \cite{Kobayashi+04, Toscani+22}: 
\begin{equation} \label{eq:tau_GW}
    f \sim \frac{1}{\tau} \approx 10^{-4}\,{\rm Hz}\times\beta^{3/2}\left(\frac{m_\star}{{\rm M}_\odot}\right)^{1/2}\left(\frac{r_\star}{{\rm R}_\odot}\right)^{-3/2}~.
\end{equation}  

Orbital parameters remain the most challenging-to-constrain variables as the TDE observables do not depend as sensitively as they do on the masses.  The $\beta$ dependency on TDE light curves is investigated with hydrodynamical simulations \cite{Guillochon+Enrico13}, resulting in co-dependency on both masses.  A relatively weaker constraint on $\beta$ is through analyzing the probability distribution among the EM-observed TDE population \cite{Wong+22}.  Both works provided corresponding analytical formulae.  For the high signal-to-noise TDE detections by \emph{LISA}, a lower bound on $\beta$ (e.g., $\beta_{\rm min}\gtrsim10$ for $M_{\rm BH}=10^6\,{\rm M}_\odot$) can be imposed \cite{Pfister+22}.  Given the rough EM-dependence, it could potentially be slightly more beneficial to further constrain $\beta$ from the TDE GW waveform as the overall shape of the polarizations and the duration of the GW burst change when $\beta$ is varied \cite{Toscani+22}.  The EM constraint can simply be used as the prior knowledge with a large uncertainty between $\beta_{\rm min} \gtrsim \beta \geq \beta_{\rm max}$, where $\beta_{\rm max}$ happens at $r_{\rm p}=r_{\rm Sch}$, i.e., all stellar materials are swallowed at pericenter passage thus EM signal detection is unlikely.  

It is important to note that these three parameters are not completely independent, e.g., a less massive, i.e., more compact, star could not be tidally disrupted by a very massive BH, but will instead be swallowed whole, i.e. $r_{\rm T}\leq r_{\rm Sch}$, where $r_{\rm Sch}$ is the Schwarzschild radius of the hole.  Some regions in the parameter space are thus automatically ruled out for any EM-observable event.

\subsubsection{Degeneracy-breaking parameters: \\
Black hole spin and inclination angle: $a_{\rm BH}$, $\theta$, $\eta$} \label{subsubsec:degparameter}

Upon first glance, the spin properties of the host BH might seem like a sub-dominant factor, but the way they correlate with the orbital inclination angle $\eta$ could ultimately lead to a probable alleviation of the known distance-inclination degeneracy.  

\begin{figure} [h!]
    \centering
    \includegraphics[width=1.00\linewidth]{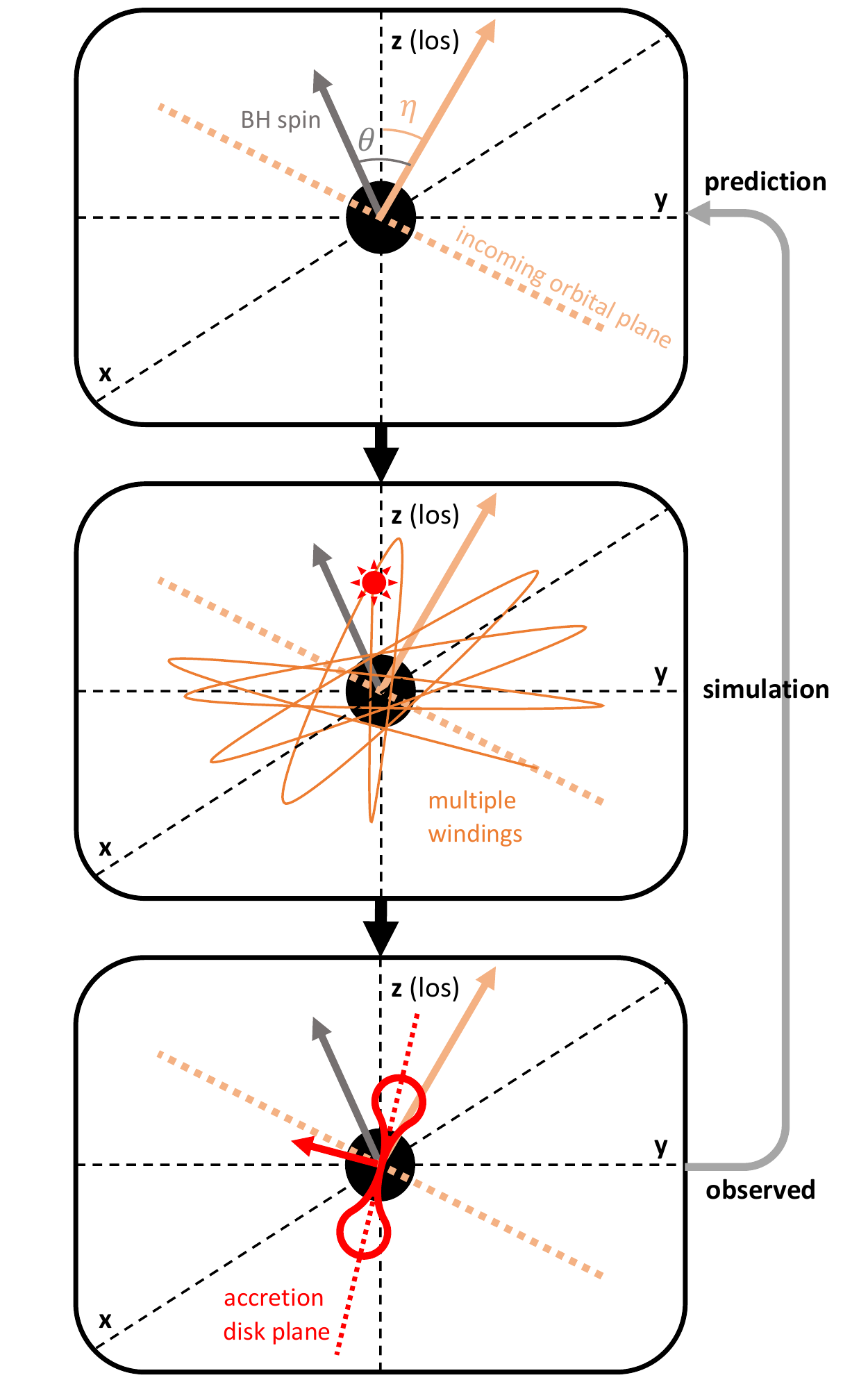}
    \caption{\textbf{Schematic illustration of how a typical $\eta$ orbit would evolve into an accretion disk of specific orientation given a BH spin offset.}  When the (large) BH spin orientation sufficiently differs from the orbital angular momentum of the stellar orbit, the disrupted stream debris would evolve away from the initial orbital plane, resulting in a stream collision somewhere else (red explosion) \cite{Guillochon+Enrico13}.  The position of the intersection could constrain the final orientation of the circularized accretion disk, where then the viewing-angle dependent model may be applied \cite{Dai+18}.  The $z$-axis is set to be the line of sight (los). \label{fig:spin-inclination}}
\end{figure}

The dimensionless spin magnitude and the spin orientation are defined by $0\leq a_{\rm BH} \equiv cJ/GM_{\rm BH}^2 \leq1$ and the angle between the spin vector and the stellar orbital angular momentum, $0^\circ\,{\rm (prograde)} \leq \theta \leq 180^\circ\,{\rm (retrograde)}$, respectively.  Spin parameters could be constrained with the observed light curve at peak accretion \cite{Kesden12} (most sensitive for high $\beta$) or with X-ray reverberation technique when the accretion disk is formed \cite{Thomsen+22a}.  If an observed TDE has a rapidly spinning host BH, the launching of a relativistic jet via the Blandford-Znajek mechanism \cite{Blandford+Znajek77} would be advantageous in further constraining the BH spin parameters \cite{Lei+Zhang11}.   

The parameter $\eta$ has never been investigated with EM observables as the incoming orbit of the star has (nearly) no influence on the multiwavelength/multi-epoch signatures as we could not see lights at the exact disruption phase.  However, when GW is included as part of the multi-messenger analysis, $\eta$ is of utmost importance.  

The first-ever analysis on incorporating viewing-angle-dependent models was used for the case of binary neutron star merger GW170817 \cite{Bulla+22}, where physical models of gamma-ray burst and kilonova are used to constrain the system inclination by modeling their light curves and spectra-photometry.  We propose that a similar approach could be implemented with the work of \cite{Dai+18,Thomsen+22b}, in which the TDE spectral features depend mainly on the viewing angle \footnote{\cite{Thomsen+22b} found the dependency between the $L_{\rm optical}/L_{\rm X-ray}$ ratio and the BH accretion rate $\dot{M}_{\rm acc}(t)$ along a specific viewing angle.  Despite the additional layer of complexity, $\dot{M}_{\rm acc}(t)$ is, however, not challenging to be deduced from observation \cite{Lodato+Rossi11}.}.  Assuming that the X-ray and optical/UV emissions originate from the inner part of the accretion disk and the disk luminosity reprocessed by the expanding outflow, respectively \footnote{The origin of optical emission is still currently under debate: some believe it is from the shocks created during the stream self-intersection \cite{Piran+15, Shiokawa+15}, while some think it comes from the reprocessing layer surrounding the disk \cite{Lodato+Rossi11, Dai+18}.}, when viewing the TDE system edge-on, the intrinsic X-ray emission from the disk will be reprocessed by the optically thick outflow, optical/UV luminosity would dominate the observed spectrum; when viewing the TDE system face-on, we would then expect to look into the optically thin funnel and see a stronger X-ray luminosity from the exposed inner disk.  Given that both optical and X-ray luminosities are measured for many TDE candidates, their ratios $L_{\rm optical}/L_{\rm X-ray}$ could potentially indicate the approximate inclination angle of the geometrically thick outer accretion disk \cite{Dai+18, Thomsen+22b}, i.e., increase in viewing angle of the disk generally augments the optical-to-X-ray luminosity ratio. 

To link the orientations of the disk and the stellar orbital plane, for BH spins that are aligned with the orbital plane's normal ($\theta=0$), we could safely assume that the stellar materials would on average remain on its incoming orbital plane due to symmetry, such that the luminosity ratio could directly be used to infer $\eta$.  For cases where the direction of the moderate/high BH spin is significantly offset from the orbital plane's normal (incoming stellar orbits are often randomly oriented), relativistic precession for close encounters (high $\beta$) would induce deflections of the debris stream out of the initial orbital plane, leading to a certain period when the TDE flare is not observable \cite{Guillochon+Enrico15}.  When the stream eventually self-intersects and dissipates its orbital energy during circularization, it is unlikely that the circularized disk will lie on the original orbital plane \cite{Tejeda+17}.  Hence, in order to utilize this analysis, both BH spin parameters should \emph{not} be neglected, their relations are illustrated by the stream evolution in Fig.\ref{fig:spin-inclination} and indicated in Fig.\ref{fig:bayesian}.  When the spin parameters are constrained by the modeling discussed in Section BH spin, the simulation \cite{Guillochon+Enrico15} could then be implemented to predict how the initial orbit plane would undergo multiple windings and end up on the final accretion disk plane where the stream intersection finally happens, hence yielding the orbital inclination angle $\eta$ through forward modeling.  

The spin parameters have to be constrained by EM observations as they are the input parameters for determining the orbital inclination angle $\eta$, meaning that if $a_{\rm BH}$ and $\theta$ are fitted through GW waveform, the distance-inclination degeneracy would still remain.

\subsubsection{Other orbital parameters: $e$, $\phi$} \label{subsubsec:otherorbit}
% may shrink this paragraph saying only 0 \lesssim e \sim 1
From the GW waveform simulation \cite{Toscani+22}, the strain amplitude dependence of the orbital eccentricity $e$ is approximately an order of magnitude smaller than $\beta$ and $\theta$.  Albeit the mild sensitivity in $e$, implying the insignificant contribution to the uncertainty of $D_{\rm L}$, EM constraints are still possible.  The common assumption is that most TDE stars have a roughly parabolic ($e\approx1$) flyby orbit \cite{Rees88}.  Hyperbolic orbits ($e\gtrsim1$) are automatically not taken into consideration since the stellar materials are unbound after the disruption and would not produce a detectable EM signal; and near-circular orbits are highly unlikely based on loss cone dynamics \cite{Merritt13}.  The eccentricity is then essentially constrained to some values in between given by the fallback rate \cite{Hayasaki+18}.  

There exists one orbital orientation parameter $\phi$, which is the angle between the stellar pericenter axis and the projection of the line of sight onto the orbital plane \cite{Toscani+22}, being the least dependent of all.  This angle can hardly be constrained by EM observations nor TDE models and shall not possess any prior in the waveform fitting.  (All angles discussed $\theta$, $\eta$, and $\phi$ are defined identically to \cite{Toscani+22}.)

\subsection{Luminosity Distance $D_{\rm L}$ and $H_0$ Estimate} \label{subsec:H0_estimate}

\subsubsection{Key Methodology} \label{subsubsec:method}
\begin{figure}
    \centering
    \includegraphics[width=1.00\linewidth]{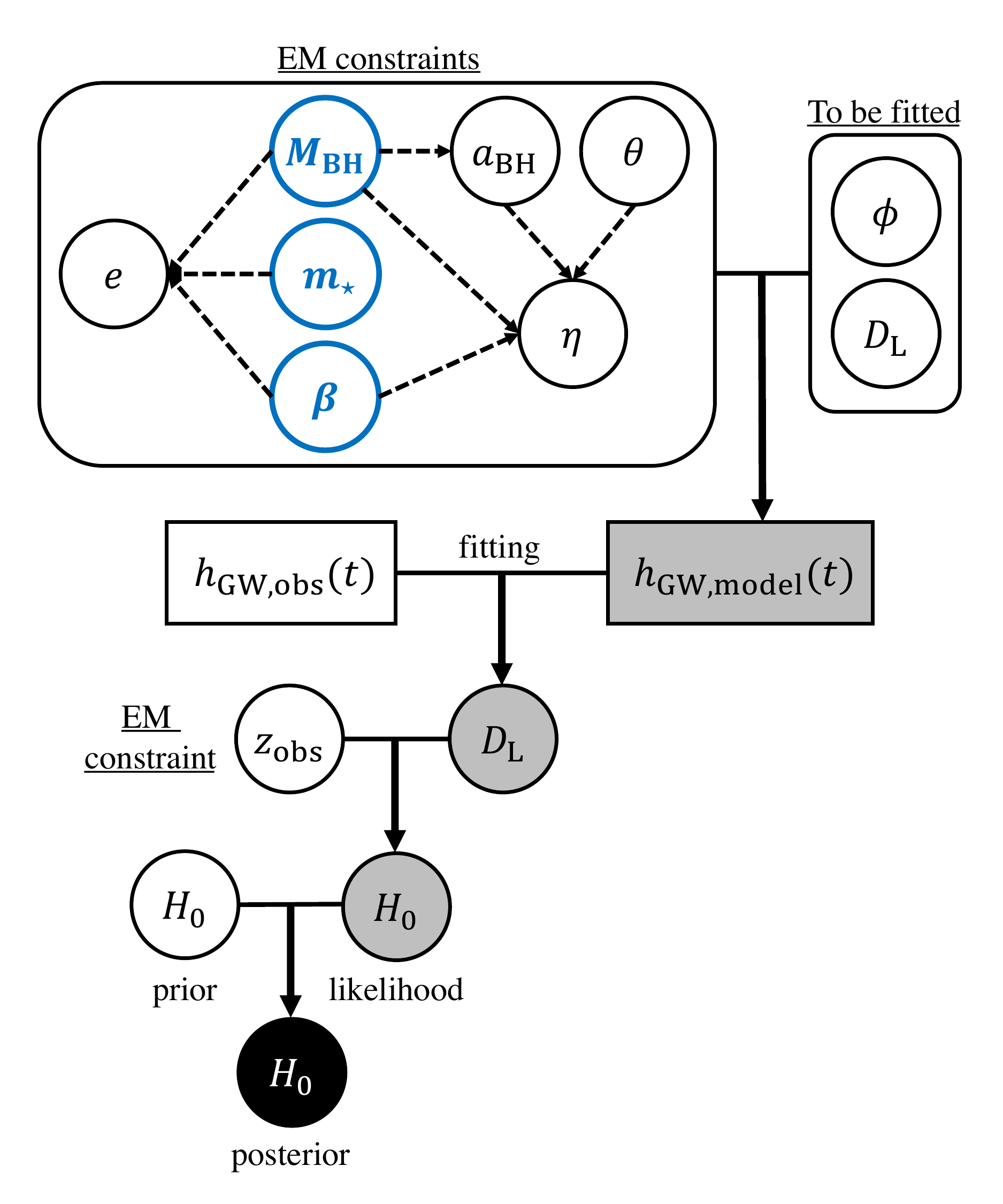}
    \caption{\textbf{Graphical model describing how the relations amongst TDE parameters and how EM and GW observations are combined to yield $H_0$ from a single TDE.}  The main TDE parameters ($M_{\rm BH}$, $m_\star$, $\beta$) are indicated by blue circles.  Dashed arrows illustrate how parameters are obtained via other parameters, which involve some simulations/models and additional EM observables (e.g., light curves and spectra).  EM-constrained parameters combining with the GW waveform fitting process would give the PDF of $D_{\rm L}$, then with the observed host galaxy redshift, the PDF of $H_0$ (likelihood) is found.  The $H_0$ likelihood and cosmologically determined prior then give rise to the posterior (the final determination of $H_0$ by a single TDE).  The filled boxes indicate the outputs of each procedure.  \label{fig:bayesian}}
\end{figure}

Using suitable TDE models to fit the corresponding observables discussed in Section \ref{subsec:EM_constraint}, all seven EM-constrained parameters ($M_{\rm BH}$, $m_\star$, $\beta$, $a_{\rm BH}$, $\theta$, $\eta$, $e$) would have their corresponding probability density functions (PDFs) obtained through simulations and model-fitting.  As for $D_{\rm L}$ and $\phi$, they are expected to freely vary during the waveform fitting, and no prior assumption should be made on $D_{\rm L}$ to prevent any bias.  With the knowledge to constrain the inclination angle $\eta$ as an input parameter, the distance-inclination degeneracy is expected to be relieved to a very large extent.  

All nine variables and their corresponding uncertainties would be used to generate a huge catalog of GW waveforms \cite{Toscani+22}, centering at the constrained parameter values to avoid exploring the large parameter space.  Note that some parameter values are prohibited ($r_{\rm T}\leq r_{\rm Sch}$), e.g., large $M_{\rm BH}$ for disruption of small $m_\star$, high $\beta$ for specific $M_{\rm BH}$ and $m_\star$.  The theoretically generated waveforms would then be fitted with that of the observed in a similar manner as in \cite{Veitch+15}, yielding a PDF of $D_{\rm L}$.  The PDF of $D_{\rm L}$ would translate directly to the PDF of $H_0$ using Eq.\ref{eq:Hubble}.  Setting this PDF as the likelihood in Bayesian formalism \cite{Abbott+17} and the $H_0$ from other cosmological studies \cite{SH0ES16, PLANCK16} as the prior, the posterior $H_0$ is expected to peak near the prior value while eliminating the other $H_0$ peaks derived from $D_{\rm L}$.  The graphical description is shown in Fig.\ref{fig:bayesian}.  

\subsubsection{Bottleneck in $H_0$ Measurement Uncertainty} \label{subsubsec:bottleneck}
\begin{figure}
    \centering
    \includegraphics[width=1.00\linewidth]{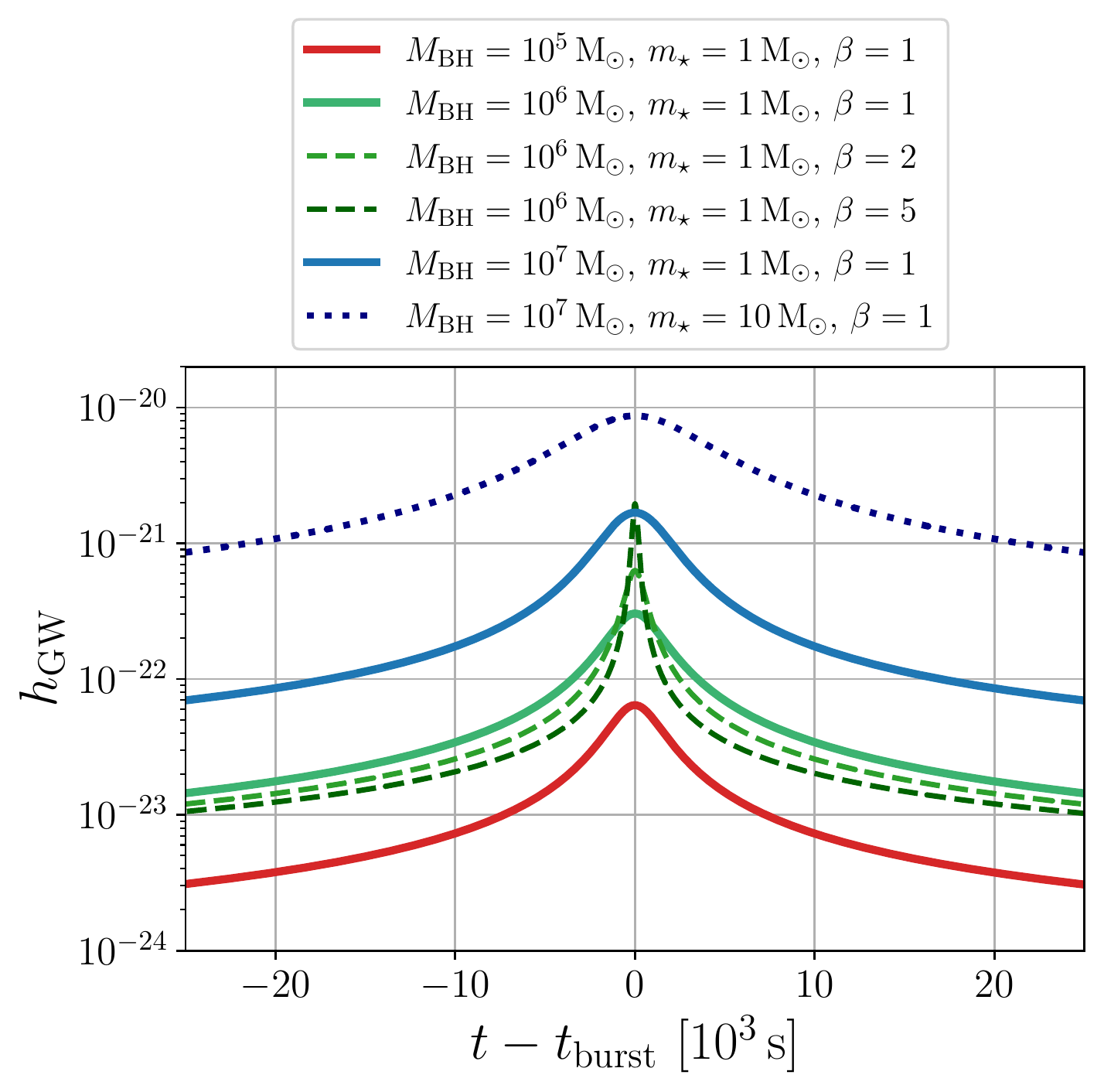}
    \caption{\textbf{TDE parameters that dominate the dependency on GW waveform amplitude $|h_{\rm GW}|$.}  Amongst the seven EM-constrained parameters, varying these three parameters: $M_{\rm BH}$ (solid), $m_\star$ (dotted), and $\beta$ (dashed), would fluctuate $h_{\rm GW}$ to a large extent.  All waveforms are centered at $t_{\rm burst}$, the time when the peak amplitude is reached.  The other parameters are chosen as follows: $a_{\rm BH}=0$, $\theta=0$, $e=1$, $\eta=0$, $D_{\rm L}=20\,{\rm Mpc}$. Plotted from simulation results \cite{Toscani+22}. \label{fig:waveform}}
\end{figure}

As long as the parameters are entangled in such a complicated manner (Fig.\ref{fig:bayesian}), what we could do is, from an order-of-magnitude point of view, estimate which parameter(s) would predominantly contribute to the uncertainty $\sigma_{D_{\rm L}}$.  

Fig.\ref{fig:waveform} shows the main TDE parameters that impact the GW burst amplitudes (the exact waveform of $h_{\rm GW}$ here is less important than those of LIGO events as TDEs often only generate single bursts), while the rest either become important only in extreme scenarios (high $\beta$ or near-maximal BH spin) or are always subdominant.  These three parameters, coincidentally, are often the input parameters in most simulations (as seen from the number of arrows pointing out of them in Fig.\ref{fig:bayesian}), thus their uncertainties are cumulative and are projected onto the rest.  Amongst them, we believe the penetration parameter $\beta$ ought to contribute the largest uncertainty of $H_0$.  Even though $M_{\rm BH}$ is used thrice to determine other parameters (while twice by $\beta$), $M_{\rm BH}$ is typically better constrained than $\beta$ \cite{Mockler+19, Ryu+20, Wong+22}, unless for very massive BHs where the detectable $\beta$ range can be as narrow as order of unity.  $\sigma_{H_0}$ is found to be dominated by the distance-inclination degeneracy \cite{Abbott+17, Bulla+22}, implying that $\sigma_{\eta}$ should dominate over the rest.  Given that $\beta$ is used to determine $\eta$, $\sigma_{\beta}$ should in turn dominate.  It is understandable as the magnitude of the off-plane precession sensitively depends on how close the stellar debris orbits around the spinning BH \cite{Guillochon+Enrico15}.  $\sigma_{M_{\rm BH}}$ would then be the next dominating uncertainty.  

The few hundred TDE GW waveforms in the presently enlarging library \cite{Toscani+22} have a resolution too low in the 9-dimensional parameter space to yield a reasonable fitting.  This should immediately raise the question: What is the approximate number of waveforms required to result in a reasonable fit, which then translates into a reasonable $H_0$ precision?  If a uniform search in parameter space is implemented, the number of waveforms generated would skyrocket as the number of parameters increase.  Having established that each parameter affects the waveform to different extents, it is only sensible to vary densely on the parameters of dominant contributions, such as $M_{\rm BH}$, $\beta$, and $\eta$.  Adaptive resolution on which parameters to explore should precede uniformly increasing the total number of waveforms across all parameter spaces in the catalog. Ultimately, the goal of the multi-messenger analysis is to better constrain $D_{\rm L}$, not finding the best-fit TDE parameters.

\section{Discussion} \label{sec:discussion}
As predicted by \cite{Kobayashi+04, Pfister+22, Toscani+22, Toscani+23}, even the optimistic TDE GW detection rate by \emph{LISA} is expected to remain a few for the entire duration of the mission.  It is therefore of utmost importance that the analyses of the few limited multi-messenger TDE observations could be maximized, stressing the power of this methodology to \emph{independently measure $H_0$ with a handful of events}.  To strengthen the constraining power of TDE parameters as a whole, the GW burst signal during disruption could trigger the immediate follow-up EM observations such that light from the pre-peak epoch can be captured.  When \emph{DECIGO} and the other next-generation spaceborne GW detectors are eventually in operation, the expected thousands to millions of TDE detections might in turn place EM observation as the bottleneck of the multimessenger era, but by then a statistically significant measurement of $H_0$ from TDEs should already be obtained.  If the uncertainty on $D_L$, hence $H_0$, could be reduced even by some small portion, after incorporating EM constraints with this proposed methodology, this would then conclusively demonstrate the functionality of TDE multi-messenger $H_0$ measurement, while placing the development of TDE modeling at the bottleneck of the analysis.  

For typical cases, $\sigma_{\beta}$ would be dominant, still, there are certain possible ways to further constrain $\beta$: by brute force, we would benefit from a GW TDE triggering of pre-peak high-cadence EM observation \cite{Guillochon+Enrico13}; more $\beta$-sensitive observable could be found with improved modeling; or exploiting the potentially huge detectable TDE population by post-\emph{LISA} interferometers, then the $\beta$-distributions \cite{Wong+22} could directly constrain $H_0$ and not the individual $\beta$ in each event.  

Given the modeling complication and intertwining relations among parameters, measuring $H_0$ with TDEs is clearly a non-trivial task and would likely require a collaborative effort in the field.  All in all, it is manifest that the proliferating EM and GW detections of TDEs and more comprehensive TDE simulations in the next decade should lead to both precise and accurate measurements of the Hubble constant in addition to the standard siren approach.

\begin{acknowledgments}
I thank S.K. Li, Paul C.W. Lai, Lars L. Thomsen, and George M. Fuller for useful comments and discussions.  
\end{acknowledgments}

%\appendix

\nocite{*}

\bibliography{apssamp}% Produces the bibliography via BibTeX.

\end{document}